\documentclass[a4paper,12pt]{article}

\usepackage{amsmath,amssymb,amsfonts}
\usepackage[dvips]{graphicx}
\usepackage{epsfig}

\usepackage{color}
\usepackage{amsmath}
\usepackage{amsfonts}
\usepackage{verbatim}
\usepackage{amssymb}
\usepackage{graphicx}
\usepackage{amsmath,amssymb,calc}
\usepackage{bbm}
\usepackage{setspace}
\usepackage{color}
\usepackage{amsmath}
\usepackage{amsfonts}
\usepackage{verbatim}
\usepackage{amssymb}
\usepackage{graphicx}
\usepackage{array}
\usepackage{epstopdf}
\usepackage{amsmath,amssymb,amsfonts,graphicx}
\usepackage{epsfig}
\usepackage[left=2.4cm,top=3.3cm,right=2.4cm,bottom=3.3cm,bindingoffset=0cm]{geometry}

\newcommand{\beq}{\begin{eqnarray}}
\newcommand{\eeq}{\end{eqnarray}}
\newcommand{\bea}{\begin{eqnarray}}
\newcommand{\eea}{\end{eqnarray}}
\newcommand{\be}{\begin{equation}}
\newcommand{\ee}{\end{equation}}

\def\brc{\langle}
\def\ckt{\rangle}

\def\1{\mathbbm{1}}

\def\nc{\sigma}
\def\ts{\widetilde{\sigma}}
\def\t{\tau}

\numberwithin{equation}{section}

\begin{document}

\title{
\begin{flushright}\ \vskip -1.5cm {\small {IFUP-TH-2016}}\end{flushright}
\vskip 20pt
\bf{ \Large Large-N CP(N-1) sigma model on a finite interval
}
\vskip 20pt}\author{
Stefano Bolognesi$^{(1,2)}$,
Kenichi Konishi$^{(1,2)}$, Keisuke Ohashi$^{(1,2,3)}$    \\[20pt]
{\em \small
$^{(1)}$Department of Physics ``E. Fermi", University of Pisa}\\[0pt]
{\em \small
Largo Pontecorvo, 3, Ed. C, 56127 Pisa, Italy}\\[3pt]
{\em \small
$^{(2)}$INFN, Sezione di Pisa,    
Largo Pontecorvo, 3, Ed. C, 56127 Pisa, Italy}\\[3pt]
{\em \small
$^{(3)}$ Osaka City University Advanced Mathematical  Institute, } \\ [3pt] 
{\em \small 
3-3-138 Sugimoto, Sumiyoshi-ku, Osaka 558-8585, Japan} \\ [3pt] 
{ \footnotesize  stefanobolo@gmail.com, }
{ \footnotesize kenichi.konishi@unipi.it,}   
{ \footnotesize   keisuke084@gmail.com}
}
\date{August  2016}
\maketitle
\vskip 0pt

\begin{abstract}

We analyze the two-dimensional $CP(N-1)$ sigma model  defined on a finite space interval $L$, with various boundary conditions, in the large $N$ limit. 
With the Dirichlet boundary condition at the both ends,  we show that the system has a unique phase, which smoothly approaches in the large $L$ limit the standard $2D$
$CP(N-1)$ sigma model in confinement phase, with a constant mass generated for the $n_i$ fields. We study the full functional saddle-point equations for finite $L$, and  
solve them numerically.  The latter reduces to the well-known gap equation in the large $L$ limit.  It is found that the solution satisfies actually both the Dirichlet and Neumann conditions. 

\end{abstract}
\newpage

\section{Introduction}

The non linear ${\mathbb C}P^{N-1}$ sigma model in two dimensions has been extensively studied in the past starting from the works \cite{D'Adda:1978uc,Witten:1978bc}. It has many properties in common with QCD  in four dimensions, such as asymptotic freedom, mass gap, confinement and the existence of a topological sector.
Moreover it has the advantage of being solvable in the large-$N$ limit.  
Recently the ${\mathbb C}P^{N-1}$ sigma model appeared  in the study of low energy effective action on certain topological solitons such as the non-Abelian string \cite{Hanany:2003hp,Auzzi:2003fs,Shifman:2004dr}.

In this paper we study  the large-$N$ solution of the bosonic ${\mathbb C}P^{N-1}$ model on a finite interval of length $L$ with various types of boundary conditions. 
The periodic boundary condition has been previously studied as a  thermal compactification in \cite{Affleck:1979gy,Actor:1985yh} and more recently in \cite{Monin:2015xwa}.   One key feature for the solvability of this model is that the periodic compactification preserves translational invariance. 

The problem with the Dirichlet boundary condition in a finite interval has been studied recently in  \cite{Milekhin:2012ca} by using a translational invariance Ansatz, where 
two possible phases were found:  a confinement phase with mass generation at large $L$ and a Higgs phase for a shorter string $L< L_{\rm crit}  \sim 1/ \Lambda$.  

It is the purpose of the present paper to examine in depth the bosonic ${\mathbb C}P^{N-1}$ model on a finite-width strip, with Dirichlet or Neumann conditions at the space boundaries. 
The exact functional saddle-point equations will be studied in the large $N$ limit, with no a priori assumption about the translational invariance. 
It will be seen that  while with the periodic boundary condition the translational invariance Ansatz is indeed consistent with the full set of equations of motion,  
 the generalized gap equations cannot be  satisfied by a translational invariant (constant) mass-generation Ansatz when the D-D or  N-N boundary condition is used. 
The solution which satisfies the full set of equations is found numerically, for various values of $L$. 

Contrary to a claim made in \cite{Milekhin:2012ca} we find that the system has a unique phase, which smoothly approaches the ``confinement phase"  in the large 
$L$ limit.  The physical reason for this result may be found in the fact that, for  small $L$,   the system reduces in the infrared (the large wavelength) limit  effectively to a one-dimensional (quantum mechanics) system where we do not expect to find any phase transitions.

\section{${\mathbb C}P^{N-1}$ sigma model on a finite space interval}

The classical  action for the ${\mathbb C}P^{N-1}$ sigma model is given by 
\beq
S= \int dx dt \left( (D_{\mu} n_i)^*D^{\mu} n_i -  \lambda (n_{i}^* n_i - r)   \right) \;,
\eeq
where $n^i$ with $i=1,\dots,N$ are $N$ complex scalar fields and the covariant derivative is $D_{\mu} =  \partial_{\mu} -i A_{\mu}$. Configurations related by a  $U(1)$ gauge transformation $z_i \to e^{i\alpha} z_i$ are  equivalent:  the $U(1)$ gauge field $A_{\mu} $ does not have a kinetic term in the classical action.
 $\lambda$ is a  Lagrange multiplier that  enforces the classical condition 
\beq
\label{classicalconstrain}
n_{i}^* n_i  = r\;,
\eeq
where  $r$ is the ``size'' of the ${\mathbb C}P^{N-1}$ manifold, which is the radius of $S^2$ sphere for $N=2$,  and  can also be expressed in terms of the coupling constant by
\beq  r= \frac{4 \pi}{g^2}\;. \eeq 
Since $n$ appears only quadratically in the Lagrangian, it can be integrated out in the partition function
\beq
Z = \int [dA_{\mu}][d \lambda][d n_i][dn_i^*] \, e^{ i S} =  \int [dA_{\mu}][d \lambda]\,  e^{ i  {S}_{{\rm eff} } }\ ,
\eeq
thus leaving an effective action for $A_{\mu}$ and $\lambda$:
\beq
{S}_{{\rm eff}} = \int d^2 x \left( N \, {\rm tr}\, {\rm log} (- D_{\mu}^*D^{\mu}  + \lambda ) -\lambda r\right) \;.
\eeq
The condition of stationarity with respect to $\lambda$ leads to the gap equation which is conveniently written  in the Euclidean formulation as
\beq
r -N \, {\rm tr} \left(\frac{1}{-\partial_{\tau}^2 -\partial_x^2   +m^2}\right) = 0\;, 
\eeq
where we have set $A_{\mu}=0$ and  $\lambda = m^2$. An expectation value of $\lambda$ provides a mass  for the $n_i$ particles (here we assumed a constant vacuum expectation value for $\lambda$; see however below).
On the infinite line the spectrum is continuous and the gap equation reads
\beq
r= N  \int_0^{\Lambda_{\rm UV}} \frac{ k dk }{2 \pi } \frac{1}{k^2 +m^2}  =  \frac{N}{4\pi} \log{\left(\frac{\Lambda_{\rm UV}^2 + m^2}{m^2}\right)} \;,
\eeq
 leading to the well-known scale-dependent renormalized coupling
\beq
\label{uvbeta}
r(\mu)= \frac{4 \pi}{g(\mu)^2}    \simeq  \frac{N}{2 \pi}  \log{\left(\frac{\mu}{\Lambda}\right)} 
\eeq
and to a dynamically generated mass, which in the case of the infinite line  can be taken to  coincide  with  the dynamical scale $\Lambda$,
\beq
\label{ml}
m = \Lambda\;. \label{eq:dynamicalscale}
\eeq

\subsection{Generalized gap equations}

The ${\mathbb C}P^{N-1}$ theory on a finite interval of length $L$,  $0 < x < L$ will now be considered.  
For this problem the boundary conditions must be specified. One possibility is the Dirichlet boundary condition which, up to a $U(N)$ transformation,  is the following constraint  on the boundary
\beq
\hbox{D-D} : \qquad n_1(0)=n_1(L) = \sqrt{r}\;,   \qquad   n_{i}(0)=  n_{i}(L) =0\;,  \quad i>1\;.  \label{DDbc}
\eeq
For simplicity and for definiteness, we take  in this paper the $n_i$ field  in the same direction in the ${\mathbb C}P^{N-1}$ space at the two boundaries. 
Another possibility is the Neumann boundary condition
\beq
\hbox{N-N}: \qquad \partial_x n_i(0) = \partial_x n_i(L) =0\;. 
\eeq
We consider in detail the three possible combinations  D-D, N-N and  periodic boundary conditions.
More possibilities are open for the choice of classical boundary condition but, for reasons to be clarified later, there is no need to list all of them.

The $N$ fields can be separated into a classical component $\nc \equiv n_1$ and the rest,   $n^i$    ($i=2,\dots,N$).
Integrating over  the $N-1$ remaining fields  yields  the following effective action
\beq  
{S}_{{\rm eff}} = \int d^2 x \left( (N-1) \, {\rm tr}\, {\rm log} (- D_{\mu}D^{\mu}  + \lambda) + (D_{\mu} \nc)^*D^{\mu} \nc -  \lambda (|\nc|^2  - r)  \right)\;.
\eeq
One can take $\nc$ real and set the gauge field to zero, and  consider  the leading contribution at large $N$ only.
The total energy  can formally be written as
\bea
\label{engen}
E =   N   \sum_{n}\omega_n + \int_0^L \left( (\partial_x \nc)^2 +\lambda (\nc^2 - r )\right) dx  \;, 
\eea
where $\omega_n^2$ are the eigenvalues of the following operator
\beq
\label{operator}
\left( - \partial_{x}^2  + \lambda(x) \right) f_n(x) = \omega_n^2 \ f_n(x)   \;,
\eeq
and $f_n$ are the corresponding   eigenfunctions.
The eigenfunctions $f_n$ can be taken to be real and orthonormal
\beq
\int_0^L   dx \, f_n(x) f_m(x) = \delta_{n\,m}\;.
\eeq
The expression for the energy (\ref{engen}) is not regularized yet, it still contains quadratic, linear and logarithmic divergences with the UV cutoff. For the moment we shall work with this formally divergent expression for the energy.
The eigenfunctions $f_n$'s satisfy also the completeness condition
  \begin{eqnarray}
\sum_{n=1}^\infty f_n(x) f_n(x')=\sum_{n\in Z}
\Big( \delta(x-x'+2\,n L) \mp  \delta(x+x'+2\,n L) \Big) \;,  \label{mirror}
\end{eqnarray}
where $\mp$ refers respectively to the Dirichlet or Neuman conditions.
Note that for $x, x^{\prime} \in [0,L]$, only the first term $\delta(x-x^{\prime})$ is relevant;  other terms guarantee that the completeness condition is consistent with 
the boundary condition (the mirror image method).

 The two functions $\nc(x)$ and $\lambda(x)$ must be determined by extremizing the energy (\ref{engen}).
Varying  the action with respect to  $\nc(x)$ one gets
\beq
\partial_x^2 \nc(x) - \lambda(x) \nc(x) = 0  \ . \label{lambda}
\eeq
In a translationally invariant system  (i.e. with constant  $\lambda$ and $\nc$) this equation reduces to $\lambda \,  \nc = 0 $. 
This  implies that either one of the condensates $\lambda$ or $\sigma$ must vanish. It follows that there are in general  two possible distinct  phases which are generally called ``confinement phase" ($\lambda \neq 0$, $\sigma =0$) or ``Higgs phase" ($\lambda =0$, $\sigma \neq0$). 
In a non-translational invariant system there is no net distinction between these two phases since in general  both $\lambda(x)$ and $\sigma(x)$ are different from zero.  

The variation of the spectrum of the operator (\ref{operator}) for $\lambda(x) \to \lambda(x) + \delta \lambda(x)$  is given by
\beq
\delta \omega_n^2 = \int_0^L dx \,  \delta \lambda(x)  f_n(x)^2 \,,
\eeq
to first order in $\delta \lambda(x)$,   as can be seen easily from (\ref{operator}). 
The variation of  the energy (\ref{engen}) with respect to $\lambda(x)$ then gives 
\beq
\label{gapeq}
\frac{N}{2} \, \sum_n\frac{f_n(x)^2}{\omega_n} + \nc(x)^2 - r = 0  \ .   
\eea
We find  it convenient  to separate  this equation into a constant  part (the average in $x$) plus a non-constant part with zero integral.
First write  $\sigma^2$  as
\beq
\sigma^2 = \ts^2 + \frac{1}{L} \int_0^L \sigma^2 dx \;, \qquad  \int_0^L \ts^2 dx  = 0\;.
\eeq
Eq.~(\ref{gapeq})  can thus be separated into the one for the constant part
\beq
\label{const}
\frac{N}{2 L} \, \sum_n\frac{1}{ \omega_n}  + \frac{1}{L} \int_0^L \sigma^2 dx  - r = 0  \;,
\eeq
and another for the non-constant part
\beq
\label{lnonconst}
\frac{N}{2} \, \sum_n\frac{1}{\omega_n}\left( f_n(x)^2 -\frac1L \right) + \ts^2 = 0 \;. 
\eeq
 So the three equations to be solved for  $\lambda(x)$ and $\sigma(x)$ are (\ref{lambda}), (\ref{const}) and (\ref{lnonconst}). Only Eq.~(\ref{const}) needs to be regularized since it contains a logarithmic divergence with the UV cutoff.

For later use we provide a relation between the expression  $\sum_n\frac{f_n(x)^2}{\omega_n}$ that appears in the generalized gap equation  and the Euclidean  propagator.
The Euclidean propagator satisfies
 \beq  
 && \left( - \frac{\partial^2}{\partial  \t^2} - \frac{\partial^2}{\partial  x^2}   +  \lambda(x) \right) D(x,\t; x',\t') = \nonumber \\
 && \quad \qquad  =    \delta(\tau-\tau^{\prime})\sum_{n\in Z}  \Big(
\delta(x-x'+2\,n L) \mp  \delta(x+x'+2\,n L) \Big)\;  \label{single} \eeq
and, in terms of $\{ f_n, \omega_n\}$, it is given by 
\begin{eqnarray}
 D(x,\tau; x',\tau')\equiv \sum_n\frac{e^{-|\tau-\tau'| \omega_n}}{2\omega_n} f_n(x) f_n(x')\;.   \label{thetwo}
\end{eqnarray}
We can thus infer the useful relation 
\beq 
\label{usefulformula}
 \sum_n\frac{f_n(x)^2}{2 \omega_n}  =   \lim_{\epsilon \to 0}  D(x, \epsilon; x ,0)\;,
\eeq
where the Euclidean time interval $\epsilon$ plays the role of the UV
cutoff.
At this stage  it is easy to see that 
the generalized gap equation (\ref{gapeq}) is nothing but 
a quantized version of the defining condition (\ref{classicalconstrain}): 
\begin{eqnarray}
\left< n_i(x)^* n^i(x)\right>-r\equiv  
\sigma(x)^2+\lim_{\epsilon \to 0}(ND(x,\epsilon,x,0)) -r=0\;.   \label{byusing}
\end{eqnarray}

\subsection{Translational invariance Ansatz \label{sec:proof}}

 Let us first test if the Ansatz of a translationally invariant  (constant) $\lambda$  is consistent with our functional saddle-point equations. 
Let us consider first the D-D case. If $\lambda= m^2$  were constant the eigenfunctions and eigenvalues of the operator (\ref{operator})  would be  simply given by 
\begin{eqnarray}
\label{ddstates}
 f_n(x)=\sqrt{\frac2L} \sin\left(\frac{n \pi x}L\right),\qquad 
\omega_n=\sqrt{\left(\frac{n\pi}L\right)^2+  m^2} \;, \qquad n\geq 1, \quad  n \in \mathbb Z\;.
\end{eqnarray}
Here one has an explicit representation of the propagator, 
\bea       D(x,\t; x',\t')  &=& \sum_{n\in Z}    \frac{1}{2\pi}  K_0 (m \sqrt{ (x- x^{\prime} + 2n L)^2 +  (\t- \t^{\prime})^2} )  \nonumber \\
&&  -    \sum_{n\in Z}    \frac{1}{2\pi}  K_0 (m \sqrt{ (x+ x^{\prime} + 2n L)^2 +  (\t- \t^{\prime})^2 )} \;.   
\eea
Note that each term satisfies the standard Green function equation (\ref{single}) with a single delta function  in $x$  on the right hand side. 
In the case with a constant $\sigma$, 
Eq.(\ref{const}) gives,  by using an appropriate regularization
\begin{eqnarray}
 0&=& \frac{N}{2L}\sum_{n}\frac1{\omega_n}+\sigma^2-r
\nonumber\\
&=& \lim_{\epsilon \to 0}\frac{1}{L} \int_0^L dx \left(
N D(x,\epsilon;x,0) -r \right)+\sigma^2\nonumber \\
&=&\frac{N}{2\pi}\ln\frac{\Lambda}m
-\frac{N}{4mL}+\frac{N}{\pi}\sum_{n=1}^\infty K_0(2nmL)+\sigma^2
\label{eq:constgap}
\end{eqnarray}
which is nothing but the gap equation 
discussed by Milekhin  \cite{Milekhin:2012ca}. Here, taking into account  the 
consistency with Eq.(\ref{eq:dynamicalscale}) for infinite string, 
the dynamical scale 
$\Lambda$ can be introduced  as 
\begin{eqnarray}
r=\frac{4\pi}{g^2}=\frac{N}{2\pi }\left(\ln\frac{2}{\Lambda \epsilon }-\gamma
	       \right)\quad \leftrightarrow \quad 
\Lambda 
=\frac{2}{\epsilon } \exp\left(-\frac{8\pi^2}{Ng^2}-\gamma\right)
\label{eq:Lambdaepsilon}
\end{eqnarray}
where $\gamma$ is the Euler-Mascheroni constant. 
The second term in the last line of Eq.~(\ref{eq:constgap})
comes from the infinite mirror poles: 
\begin{eqnarray}
\frac{N}{2\pi} \frac{1}{L} 
\sum_{n \in Z}  \int_0^L  dx  K_0 (2m | x + n L|) 
=    \frac{N}{2\pi} \frac{1}{L} \int_{-\infty}^{\infty}  dx \,  K_0 (2m
| x |) =    \frac{N}{4 m L}\;. \label{eq:mirrors}
\end{eqnarray}

One can verify that 
Eq.~(\ref{eq:constgap}) gives the local extremum (maximum) 
of the total energy (\ref{engen}) 
\begin{eqnarray}
 E&=& \lim_{\epsilon \to 0} \int_0^L  \left\{ 2 \frac{\partial^2}{\partial
				       \epsilon^2}(N D(x,\epsilon;x,0
				       )-r) 
+m^2 (\sigma^2-r)\right\}dx\nonumber \\
&=&  \frac{Nm^2L}{2\pi} \left(\frac12 -  \log \frac m{\Lambda} \right)-
 \frac{N m}2 
-\frac{ N m^2 L}\pi \sum_{n=1}^\infty \frac{ K_1(2 \, n \,m\,L)}{n\,m
L}+m^2L \sigma^2       \label{energycstm}
\end{eqnarray}
with respect to $m^2$,   under the translational-invariance Ansatz.

However,  the problem is that an additional equation, (\ref{lnonconst}),  must be satisfied as well. 
The left hand side, using  $\ts = 0$ (under the assumption of translational invariance), is  just
\begin{eqnarray}
\label{DDnosolved}
 \frac{N}{2L} \sum_{n=1}^\infty \frac1{\omega_n}  \left( 2 \sin^2
						   \left(\frac{n \pi
						    x}L\right)  - 1
						  \right) 
=-\frac{N}{2L} \sum_{n=1}^\infty \frac1{\omega_n}  \cos \left(\frac{2 n \pi x}L\right) \;,
\label{formuladd}
\end{eqnarray}
which  is equal to, by using (\ref{usefulformula}), 
\bea   
&&  N\lim_{\epsilon \to 0} \left(  D(x, \epsilon; x;0)-  
 \frac{1}{L}  \int_0^L   dx  D(x, \epsilon; x;0)\right)  
\nonumber\\
&=& 
\frac{N}{2L}  \left( \frac1{2 m }-\frac{L}\pi \sum_{n  \in \mathbb Z}K_0(2  m  |x- n L|) \right).\label{propeq}
\eea
This is clearly not zero:   Eq.~(\ref{lnonconst}) is not
satisfied by a  constant $m$.  

 For N-N boundary conditions and   $\lambda=m^2$ constant the eigenmodes  are
\begin{eqnarray}
\label{nnstates}
 f_n(x)=\sqrt{\frac2L} \cos \left(\frac{n \pi x}L\right),\quad 
\omega_n=\sqrt{\left(\frac{n\pi}L\right)^2+ m^2} \;, \qquad n\geq 0\,, \quad n \in \mathbb Z
\end{eqnarray}
In this case,  the left hand side of Eq.~(\ref{lnonconst}) gives 
\begin{eqnarray}
\label{NNnosolved}
 \frac{N}{2L} \sum_{n=0}^\infty \frac1{\omega_n}  \left( 2 \cos^2 \left(\frac{n \pi x}L\right)  - 1 \right)  
= 
\frac{N}{2L} \sum_{n=1}^\infty \frac1{\omega_n}  \cos \left(\frac{2 n \pi x}L\right)\nonumber \\
=-\frac{N}{2L}  \left( \frac1{2 m }-\frac{L}\pi \sum_{n  \in \mathbb Z}K_0(2 m |x-n L|) \right).
\end{eqnarray}
which is the same as  the D-D case (\ref{DDnosolved}), except for the sign.
The proof is as before by using instead the plus sign in (\ref{single}). 
Again, Eq.~(\ref{lnonconst}) cannot be satisfied by a constant $\lambda$.

Thus both for DD and NN boundary conditions, the constant generated mass $m$ does not represent  a quantum saddle point of our system. 

For the periodic boundary condition one can set for convenience the period length to $2L$. The complete set of eigenstates and eigenfunctions are exactly given by a D-D set  (\ref{ddstates}) plus a N-N set, (\ref{nnstates}),  applied in the enlarged space $[0, 2L]$.
In this case,  the left hand side of equation (\ref{lnonconst}) is the sum of D-D states contribution (\ref{DDnosolved}) plus the N-N states contribution (\ref{NNnosolved}) which cancel.   As expected,  the translational invariance Ansatz is consistent in this case. 
This system has been studied in \cite{Monin:2015xwa}. 
With a periodic condition, 
the total 
energy and the gap equation can depend on the Wilson loop
$\exp(i \int_0^{2L}A_xdx)$.  With a nonvanishing constant gauge field $A_x$,  one finds that 
 the propagator is  given by
\begin{eqnarray}
 D(x,\t; x',\t')  &=& \sum_{n\in Z}    \frac{1}{2\pi}  K_0 (m \sqrt{ (x- x^{\prime} + 2n L)^2 +  (\t- \t^{\prime})^2} ) \, e^{iA_x(x-x'+2nL)}.
\end{eqnarray}
Therefore the energy in the confinement phase $(\sigma=0)$,  is 
\cite{Monin:2015xwa} 
\begin{eqnarray}
 E&=&2L\times \lim_{\epsilon\to 0}\left\{
2\frac{d^2}{d\epsilon^2}(ND(0,\epsilon,0,0)-r)-m^2 r\right\}\nonumber\\
&=&
\frac{Nm^2L}{\pi} \left(\frac12 -  \log \frac m{\Lambda} \right)
-\frac{2 N m^2 L}\pi \sum_{n=1}^\infty \frac{ K_1(2n  \,mL)}{n\,m
L}\cos(2nA_x L).
\end{eqnarray}
By minimizing the energy with respect to $A_x$,  one finds $A_x=0$. 
The saddle point with respect to $m^2$ is determined by the gap equation    
\begin{eqnarray}
0= \frac{N}{2\pi}\ln\frac{\Lambda}m+\frac{N}{\pi}\sum_{n=1}^\infty K_0(2nmL).
\end{eqnarray}
In this case no other equations must be satisfied,  so this gives the solution of the problem.

\section{Solution of the generalized gap equation}

Having  proven that a constant $\lambda=m^2$ does not satisfy the functional saddle point equations, (\ref{lambda}), (\ref{const}) and (\ref{lnonconst}), for the problem with the Dirichlet or Neumann boundary conditions,  
   we now discuss the solution of these equations, relying also on a numerical analysis.   The content of this section constitutes the main new contribution of this paper.

\subsection{Behavior of the functions $\lambda(x)$ and $\sigma(x)$ near the boundaries}

Before embarking on the numerical analysis  one must  first understand  the behavior close to the boundaries of the two unknown functions $\lambda(x)$ and $\sigma(x)$. 
For the case of D-D boundary condition, and under the assumption of constant $\lambda$,  the left hand side of  (\ref{lnonconst}) is given by  (\ref{DDnosolved}) and thus contains a logarithmic divergence near the boundaries since $K_0(x) \simeq -\log{x}$ for $x \ll 1$.
In the propagator formalism (\ref{propeq}) this leading divergence comes from the closest mirror pole. 
The logarithmic behavior is due to the fact that at $x \ll 1$ one can neglect the potential $\lambda$ and use the massless propagator.
Another crude, but somehow useful, way to understand this logarithm divergence is the following. At fixed $x$, we split the sum (\ref{formuladd}) into ``low-energy" modes and ``high-energy" modes. 
For the low-energy modes we approximate $ \sin^2 \left(\frac{n \pi x}L\right)  \simeq 0$. The high-energy modes instead are summed to zero due to the fast averaging. The result gives the correct  log divergent term:
\beq
 && \frac{N}{2L} \sum_{n=1}^{[L/\pi x]} \frac1{\omega_n}  \left( 2 \sin^2 \left(\frac{n \pi x}L\right)  - 1 \right) +
 \frac{N}{2L} \sum_{n=[L/\pi x]+1}^{\infty} \frac1{\omega_n}  \left( 2 \sin^2 \left(\frac{n \pi x}L\right)  - 1 \right) \nonumber \\
&& \simeq 
 \frac{N}{2L} \sum_{n=1}^{[L/\pi x]} \frac1{\omega_n}  \left(- 1 \right) \simeq  \frac{N}{2 \pi} \log{x} + \dots \ .
\end{eqnarray}

We will show below more carefully that this type of divergence near the boundary is really there, at least for the D-D condition,  also for a generic, but not too singular, potential $\lambda(x)$. 
Let us  use the WKB approximation to compute the leading divergence of $ \sum_n\frac{f_n(x)^2}{\omega_n}$ near the boundaries of the interval. We assume  $\lambda(x)$ to be generic, with $
\lambda(x) >0$ and $ \lambda(x)=\lambda(L-x) $,  but not too singular near the boundaries:
\beq
\label{condition}
\lim_{x \to 0 } x^2 \lambda(x) = 0\;.
 \eeq
It will be seen that the potential $\lambda(x)$  is non-constant and
divergent near the boundaries $\lambda(x) \to \infty$ as $x \to
0,L$. Nevertheless it does satisfy the condition (\ref{condition}). 
This fact also implies that `the Higgs phase' defined by a 
configulation in which the potential vanishes identically, $\lambda(x)=0$,
cannot be a solution of the generalized gap equation. 
By using WKB approximation  the high-energy modes of the operator (\ref{operator}) can be approximated, in the classically allowed region, by
\begin{eqnarray}
 f_n(x) \simeq \frac{C_n}{\sqrt{p_n(x)}} 
\sin\left(\int_{\epsilon_n}^x p_n(x') dx'-\frac{\theta_n}2\right)
\end{eqnarray}
where 
\begin{eqnarray}
p_n(x)=\sqrt{\omega_n^2-\lambda(x)},\qquad C_n =  
 \sqrt{\frac2{\int_{\epsilon_n}^{L-\epsilon_n} \frac{dx}{p_n(x)} }}
\end{eqnarray}
and $\epsilon_n$ and $L-\epsilon_n$  are the classical turning point
\beq
\omega_n^2-\lambda(\epsilon_n) = 0 \;,\qquad   \omega_n^2-\lambda(L-\epsilon_n) = 0 \ .\label{eq:turnigpoint}
\eeq
The Bohr-Sommerfeld quantization condition is 
\begin{eqnarray}
 \int_{\epsilon_n}^{L-\epsilon_n} \sqrt{\omega_n^2-\lambda(x)} \, dx 
=\pi \, n+\theta_n\;,\qquad n\in\mathbb Z_{>0}\;.
\end{eqnarray}
The initial phase $\theta_n$ is quite important to determine the divergence
near the boundaries.   It would be equal to ${\pi}/2$ if there were no boundary, but a smooth classical turning point.
However, the phase $\theta_n$ turns out to  vanish 
\begin{eqnarray}
 \theta_n=0\;,
\end{eqnarray}  
as can be seen as follows. If $\lambda(x)$ had the maximum $\lambda_M$, 
then for $\omega_n >\sqrt{\lambda_M}$ there was no classical turning
point and the classically allowed
region was $[0,L]$   $(\epsilon_n=0)$  and we get $\theta_n=0$ due to
the D-D condition.  Even in the case where $\lambda$ diverges at the boundaries, which
actually is our case,   the same result is obtained because, from the condition (\ref{condition}) and the equation
(\ref{eq:turnigpoint}),  one gets
\beq
\lim_{\omega_n \to \infty} \omega_n \, \epsilon_n = 0 \;,
\eeq
meaning that  the turning point $\epsilon_n$ goes to zero faster than the typical wave length of the eigenstate, $\sim 1/\omega_n$.
This is sufficient for letting us  to conclude that for large $n$ the eigenfunctions and eigenvalues are well approximated by
\beq
\label{largen}
f_n \simeq 
\sqrt{\frac2L} \sin\left(\frac{n \pi x}L\right)\;,  \qquad  
\omega_n \simeq \frac{n\pi}L\;,
\eeq
and thus 
\beq
\frac{N}{2} \, \sum_n\frac{1}{\omega_n}\left( f_n(x)^2 -\frac1L \right)  \simeq-\frac{N}{2 \pi } \sum_{n=1}^\infty \frac1n  \cos \left(\frac{2 n \pi x}L\right)
= \frac{N}{2 \pi} \log
\left(2 \sin{\left(\frac{\pi x}{L} \right)}  \right) \ ,  \label{massless}
\eeq
which behaves indeed  near $x=0$   as 
\be    \sim    \frac{N}{2 \pi}   \log x\;.
\ee

In other words,  the sum of the fluctuations of the $n_i$ fields very generally produces  a divergent behavior for the first two terms of  Eq.~(\ref{lnonconst})   near the boundaries, 
 quite independently of the form of $\lambda(x)$, as long as   the latter satisfies the condition (\ref{condition}). 
The only way to solve  Eq.~(\ref{lnonconst}) is then to cancel this divergence with the classical field $\sigma$
\beq
\label{sigmadiv}
 \sigma^2 \simeq   \frac{N}{2 \pi}  \log{\frac{1}{x}}   \;. 
\eeq
But this is exactly the expected behavior, as  it corresponds to  the classical constraint (\ref{classicalconstrain}),  or  to  the D-D boundary condition, (\ref{DDbc}),  as can be seen once   a renormalized coupling (\ref{uvbeta}) with energy scale $\mu =1/x$ is substituted. Approaching the boundaries, $x \sim \epsilon,   L- \epsilon $,  is thus similar to  going into the UV, hence the classical value for $\sigma(x)=n_1(x)$ there (see more on this point below).

The divergence of the $\sigma$ field near the boundaries (\ref{sigmadiv}) on the other hand  implies, through  equation (\ref{lambda}), that  the potential $\lambda$ also diverges. Its leading behavior  at $x \simeq 0$ is given by 
\beq
\lambda(x) \simeq    \frac{1}{2  \,x^2 \log{1/x}}  \; \label{satisfying}
\eeq
which  indeed satisfies the condition, (\ref{condition}).   

Thus the behavior of the functions $\lambda(x)$ and $\sigma(x)$ near the boundaries has been determined self-consistently  to be given by (\ref{sigmadiv}) and (\ref{satisfying})
(and similarly for $x\sim L$, with $x \to L-x$). 

Physically, such a behavior can be understood as follows.  At any fixed $x$, far from the boundaries, the UV divergences due to the higher modes are cancelled by a constant 
(and divergent)  $r_{\epsilon} = \tfrac{4\pi}{g_{\epsilon}^2} $, exactly as in the standard  ${\mathbb C}P^{N-1}$ sigma model without the space boundaries. This is so because the UV divergences are 
local effects in $x$:  far from the boundaries the effects of the latter are negligible. Near the boundaries where the quantum fluctuations of the $n_i$ fields are suppressed, however, this cancellation cannot occur,  $r_{\epsilon} $  being  a constant. This is where the classical field $\sigma= n_1$ comes to rescue. 
The generalized gap equation (\ref{gapeq}) is now  satisfied by balancing the "ultraviolet" divergent behavior  of $\sigma(x)$ at $x \sim \epsilon  $,   with $r_{\epsilon}$. 

Even though this way of  interpreting  (\ref{sigmadiv}) and (\ref{satisfying}) is straightforward in the case of the D-D condition, actually the same 
results hold for the system with the N-N boundary condition as well.  They just tell that at the boundaries,  where the quantum fluctuations  of the  $n_i$ fields are suppressed and cannot produce the logarithmic divergences due to the lack of the two-dimensional spacetime,  the defining condition (\ref{byusing})  of the  ${\mathbb C}P^{N-1}$ theory  must be satisfied by the classical field \footnote{Note that  in the large $N$ approximation we are working in, ``quantum fluctuations" of the $\sigma$ field are  suppressed by $1/N$ as compared to those of $n_i, \, i\ne 1$, as well as  to its classical mode.},  independently of the type of the boundary condition.
The numerical solutions found below indeed hold true both for the case of the D-D and the N-N  boundary conditions.  

This conclusion can be further confirmed by studying  the wave functions $f_n(x)$ analytically near the boundaries,  assuming  $\lambda(x)$ of the form of  (\ref{satisfying}), as is done in
  Appendix \ref{eigen}.   The resulting behavior is  
\beq
f_n(x \to 0) \propto \frac{x}{\sqrt{-\log{x}}}\;;  \qquad 
f_n(x \to L) \propto   \frac{L-x}{\sqrt{-\log{(L-x)}}}\;.     \label{result}
\eeq
Thus not only do they satisfy the Dirichlet boundary condition but  also the Neumann boundary condition.
As already anticipated,  this is the indication one gets  from the numerical solution found by the recursive procedure to be described in the next subsection. The results found for $\lambda(x)$ and $\sigma(x)$ 
turn out to be the same, irrespectively of the boundary condition used to solve the wave equation (\ref{operator}). 

Finally we note that  the case of the periodic boundary condition is qualitatively different. In that case, the points $x=0, L$ are no special points, the quantum fluctuations 
are effective there, as well as at  any other point, and our results (\ref{sigmadiv}) and (\ref{satisfying}) do not apply.  Such system has been recently studied in \cite{Monin:2015xwa}.

\subsection{Numerical solutions for $\lambda(x)$ and $\sigma(x)$}

The method we employ to solve the set of equations
 (\ref{lambda}), (\ref{const}), and (\ref{lnonconst}) numerically is as follows. 
 We first introduce a  finite cutoff $n_{\rm max}$ on the number of modes. Accordingly,  equation (\ref{const}) can be regularized as  (see Eq.~(\ref{uvbeta}))
 \footnote{
A careful change of the regularization gives 
the factor 2 in front of $\mu(n_{\rm max})$ 
so that it is consistent with Eq.(\ref{eq:dynamicalscale}).
Omitting higher modes larger than $n_{\rm max}$ 
\begin{eqnarray}
 \frac{\pi}{L}\sum_{n=n_{\rm max}+1}^\infty \frac{e^{-\epsilon
  \omega_n}}{\omega_n}\simeq \sum_{n=n_{\rm max}+1}^\infty 
\frac{e^{-\frac{n \pi}{L}\epsilon}}{n}
\simeq -\log(1-e^{-\frac{\pi}L\epsilon })-\sum_{n=1}^{n_{\rm
max}}\frac1n
\simeq -\log \epsilon -\log \mu(n_{\rm max})-\gamma\;,
\end{eqnarray}
introduces  a translation between the two UV cut-off parameters
 $\epsilon$ and $\mu(n_{\rm max})$ as $1/\epsilon =\mu
 e^{\gamma}$. Therefore, Eq.(\ref{eq:Lambdaepsilon})
giving $\Lambda$ is translated as 
\begin{eqnarray}
 r=\frac{N}{2\pi}\left(\log\frac2{\Lambda \epsilon }-\gamma
		 \right)=\frac{N}{2\pi} \log \frac{2\mu(n_{\rm max})}{\Lambda}\;.
\end{eqnarray} 
} 

\beq
\label{lconstreg}
\frac{N}{2 L} \, \sum_{n=1}^{n_{\rm max}}\frac{1}{ \omega_n}  + \frac{1}{L} \int_0^L \sigma^2 dx  - \frac{N}{2 \pi}  \log{\left(\frac{2\mu(n_{\rm max})}{\Lambda}\right)}  = 0 \;,
\eeq
where, by using (\ref{largen}),
\beq
\mu(n_{\rm max}) = \frac{n_{\rm max} \pi}{L} \ .
\eeq
The algorithm we use  here to find the solution,  somewhat reminiscent of Hartree's equations  in atomic physics,  is a recursive procedure  in which at each step $k$ one has  a certain function $\lambda_k(x)$, with some initial $\lambda_0(x)$ satisfying (\ref{satisfying}).  
 Given $\lambda_k(x)$, the Schr\"odinger equation (\ref{operator}) is  solved to give $ \{ \omega_n,  f_n(x)\}$ for the desired number of modes;    (\ref{lnonconst}) and  (\ref{lconstreg})  are then used  to find the corresponding value of $\sigma_k^2$:
\beq
\label{sigmak}
\sigma_k^2 = -
\frac{N}{2} \, \sum_n\frac{1}{\omega_n}\left( f_{n,k}(x)^2 -\frac1L \right)-
\frac{N}{2 L} \, \sum_{n=1}^{n_{\rm max}}\frac{1}{ \omega_{n,k}}
 +\frac{N}{2 \pi}  \log{\left(\frac{2 \mu(n_{\rm max})}{\Lambda}\right)} \;.
\eeq
Note that the right hand side of this equation depends only on $\lambda_k(x)$.

The next  iteration for $\lambda$ is found by use of  Eq.~(\ref{lambda}):
\beq
\label{weakpassage}
 \lambda_{k+1}(x)  = \frac{\partial_x^2 \nc_k(x)}{\nc_k(x) }  \ .
\eeq
This procedure can be  repeated  until a convergence to a consistent
set of   $\lambda(x)$ and $\sigma(x)$ is achieved\footnote{The whole
analysis has been performed  by using Mathematica, Wolfram (Version 10.3).   The
spectra $\{f_n, \omega_n\}$ at each recursion step are found by use of
the command ``NDEigensystem''.
}.
We set $L$ fixed and then study various values of $\Lambda$.  We take the starting point of the iteration to be  $\lambda_0(x)=0$ for the first value of $\Lambda$.  
For the successive values of $\Lambda$,   the self-consistent solution  $\lambda(x)$ obtained with the previous value of $\Lambda$ will be taken as the 
starting point $\lambda_0(x)$ of recursion.  

The method as described above works well for $L\Lambda$ up to $1$ after which it loses convergence.  
The delicate  point in the algorithm is the step (\ref{weakpassage}) which is exceedingly  sensitive to small variations of $\sigma_k(x)$  in the denominator. 
A technical improvement we implement is the following. We substitute (\ref{weakpassage}) with  a ``smoother" recursive formula   
\beq
\label{strongpassage}
 \lambda_{k+1}(x)  = \frac{1}{K} \left( \frac{\partial_x^2 \nc_k(x)}{\nc_k(x) } + (K-1) \lambda_k(x)\right) \ ,
\eeq
with some conveniently chosen $K$. Having a larger value of $K$ makes the convergence slower but at the same time it allows to reach higher values of $L \Lambda$. For example by choosing $K= 15 $ (as is done to produce the plots for large $L \Lambda$  given below) one  can reach $L \Lambda $ up to  $4$.

Another technical issue is that  $\sigma_k(x)$, as computed in (\ref{sigmak}), cannot be inserted as it is in the second derivative of the expression  for $\lambda_{k+1}(x)$.  In passing from  (\ref{sigmak})  to  (\ref{strongpassage}) one must  perform a numerical fit to approximate   $\sigma_k(x)$  with a sum of analytical functions. In the plots given below  $\sigma_k(x)$ and its analytical fit are superimposed: the two curves are indistinguishable, within the resolution of the graphs.

\noindent {\bf Results} 

\begin{figure}[h!t]
\centerline{
\begin{tabular}{cc}
\epsfxsize=8cm \epsfbox{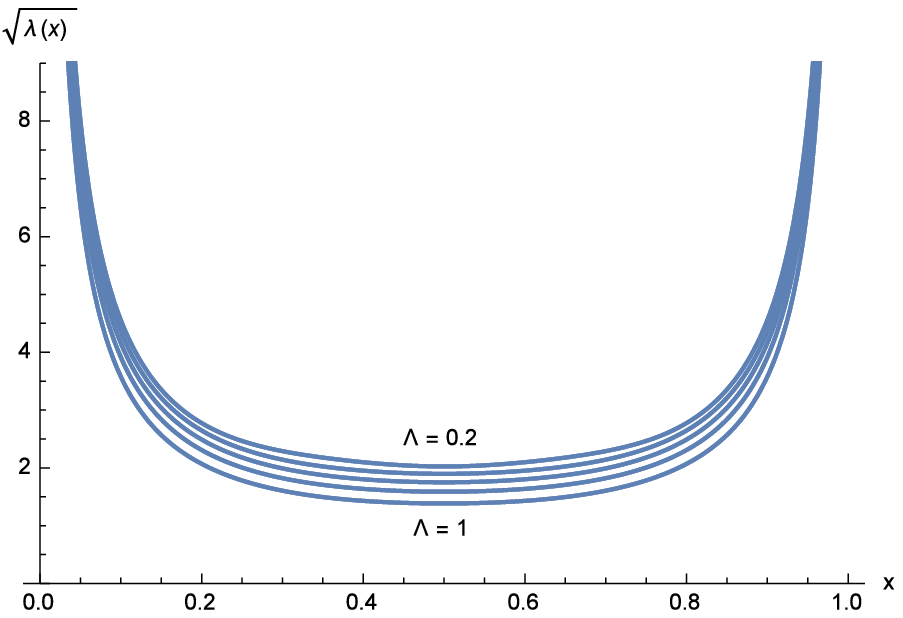}   & 
\epsfxsize=8cm \epsfbox{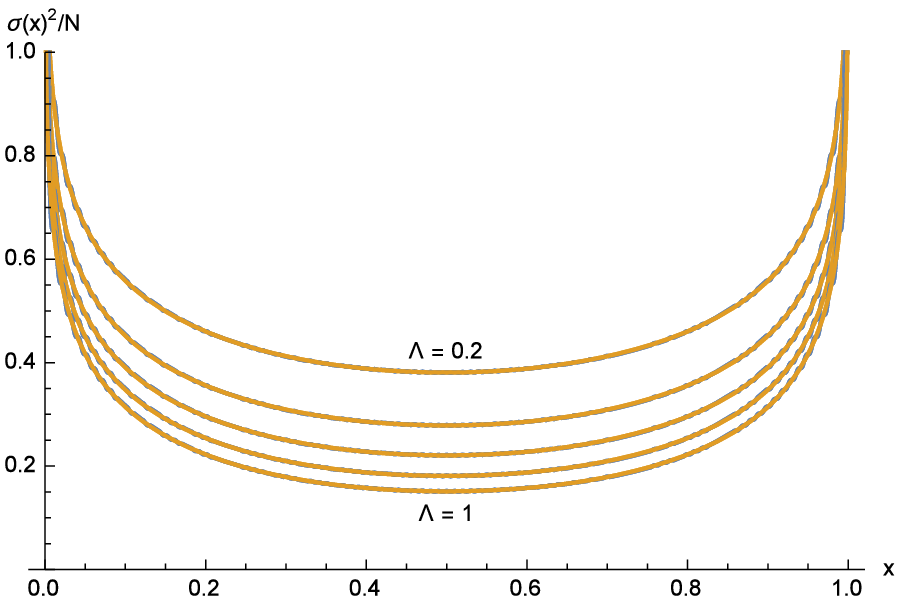} 
\end{tabular}}
\caption{{\footnotesize On the left $\sqrt{\lambda(x)}$, on the right $\sigma(x)^2/N$. These plots are obtained by keeping $L=1$ fixed and changing $\Lambda$. }}
\label{lconst}
\end{figure}

In Figure \ref{lconst} we present some numerical results for smaller values of $L \Lambda$.   $\sqrt{\lambda(x)}$  is shown in the
left panel  while  the right panel  gives  the classical field $\sigma(x)^2/N$.
The Figures show the results for $L=1$ fixed and $\Lambda = 0.2, 0.4, 0.6, 0.8, 1$.   

\begin{figure}[h!t]
\centerline{
\begin{tabular}{cc}
\epsfxsize=8cm \epsfbox{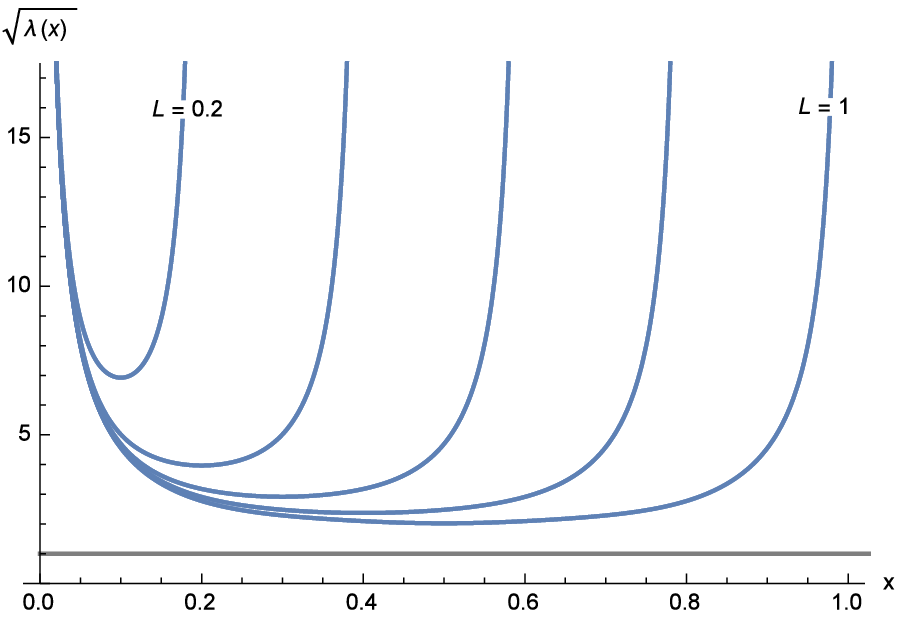}  & 
\epsfxsize=8cm \epsfbox{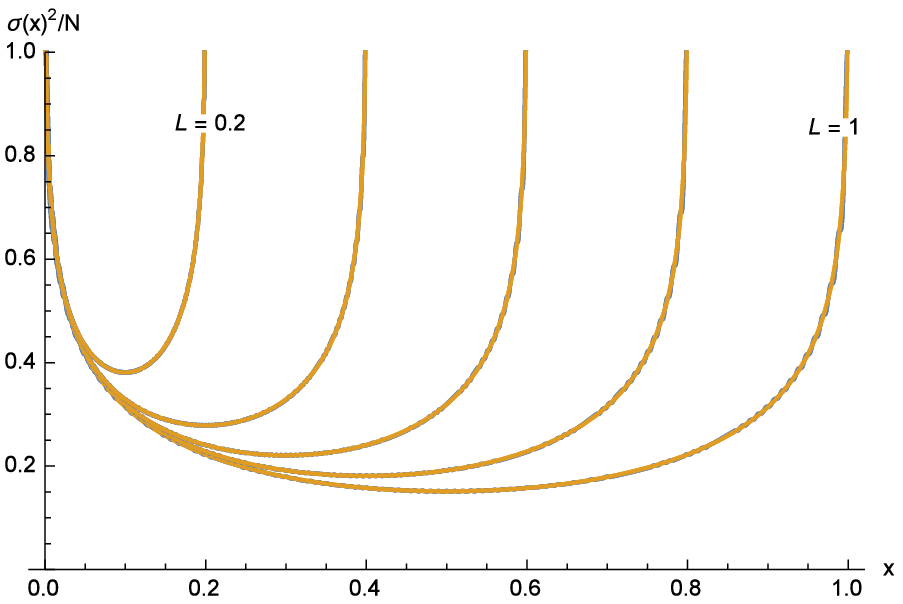} 
\end{tabular}}
\caption{{\footnotesize On the left $\sqrt{\lambda(x)}$, on the right $\sigma(x)^2/N$. These plots are the corresponding of Figure \ref{lconst} but rescaled in order to keep $\Lambda=1$ fixed and changing $L$. The constant line on the left figure is $\Lambda$. }}
\label{lambdaconst}
\end{figure}
Figure \ref{lambdaconst} shows the plots for fixed dynamical scale $\Lambda = 1$ and various values of $L$. 
The results in the two
figures are related by a trivial rescaling. The only non-trivial dimensionless quantity is $ \Lambda L$. Two cases with the same $ \Lambda L$ are related by the following scale transformation
\bea
\frac{1}{\alpha^2}\lambda\left(\frac{x}{\alpha};\frac{L}{\alpha},\Lambda \alpha\right) =  \lambda(x;L,\Lambda)\;; \qquad 
\sigma\left(\frac{x}{\alpha};\frac{L}{\alpha},\Lambda \alpha\right) = \sigma(x;L,\Lambda)\;.
\eeq
\begin{figure}[h!t]
\centerline{
\begin{tabular}{cc}
\epsfxsize=8cm \epsfbox{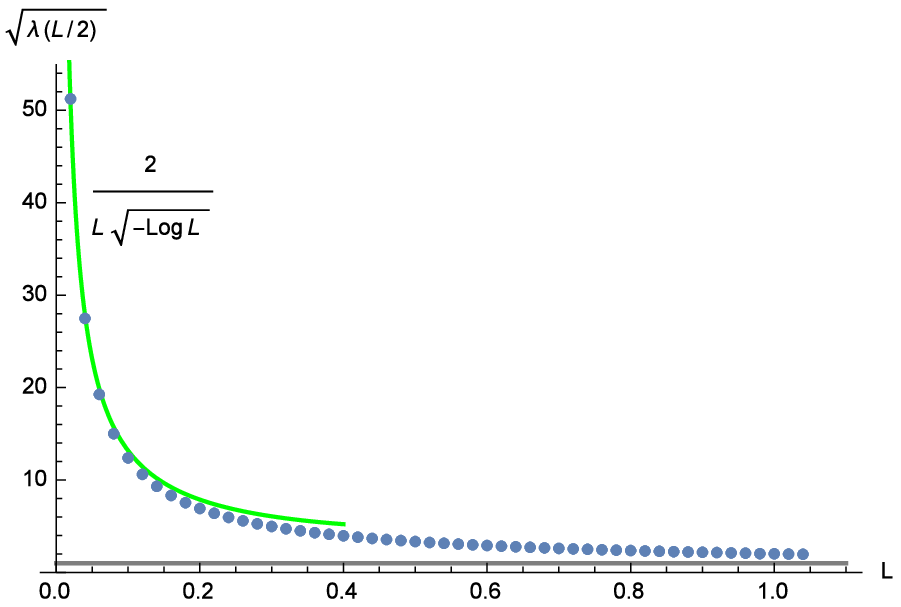}  & 
\epsfxsize=8cm \epsfbox{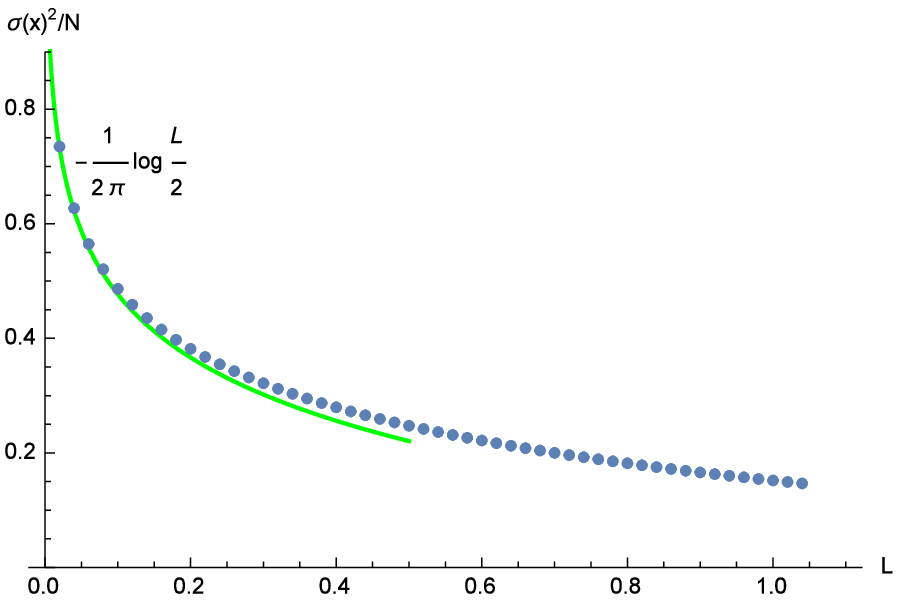} 
\end{tabular}}
\caption{{\footnotesize On the left  is $\sqrt{\lambda(L/2)}$;  on the right $\sigma(L/2)^2/N$ for various values of $L$ keeping $\Lambda=1$. The data points are compared with the singular behavior (\ref{sigmadiv}) and (\ref{satisfying}) extrapolated towards the center of the string, $L/2$.   }}
\label{mezzo}
\end{figure}
Another visualization of the results is given in Figure \ref{mezzo} where the value of the fields at the center of the string, $L/2$, is shown for various values of $L$ and at fixed $\Lambda =1$. 

\begin{figure}[h!t]
\centerline{
\begin{tabular}{cc}
\epsfxsize=8cm \epsfbox{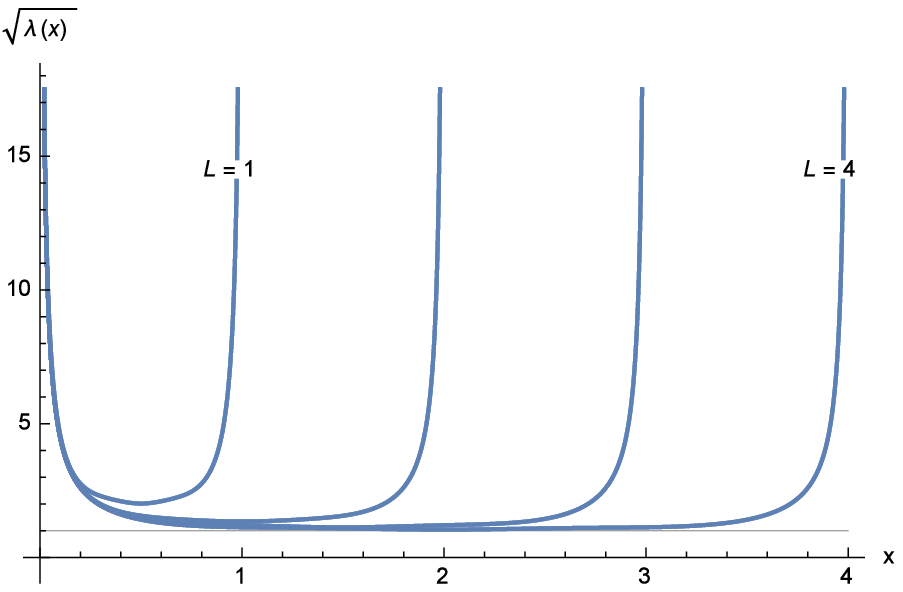}  & 
\epsfxsize=8cm \epsfbox{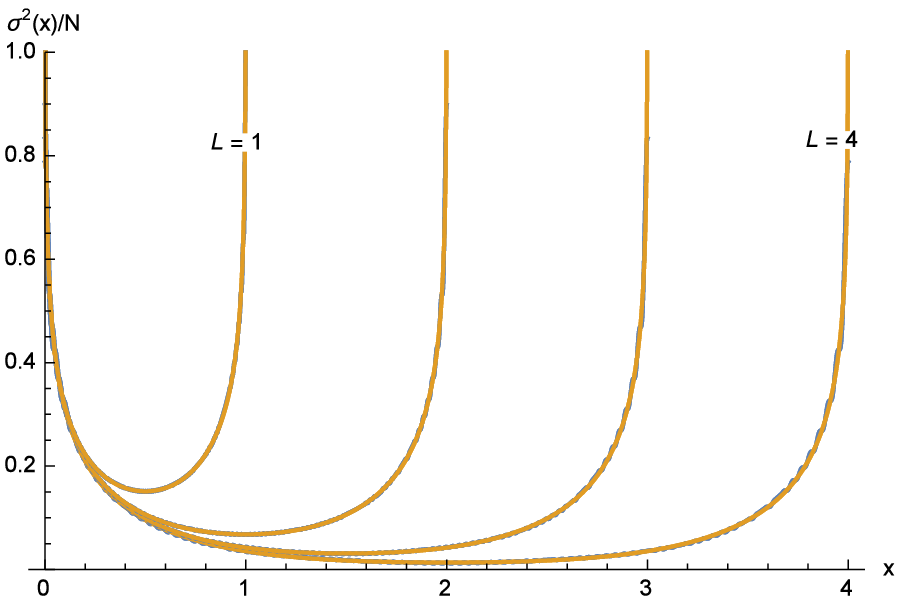} 
\end{tabular}}or 
\caption{{\footnotesize On the left $\sqrt{\lambda(x)}$, on the right $\sigma(x)^2/N$. These plots are rescaled in order to keep $\Lambda=1$ fixed and for  $L=1,2,3,4$. The constant line on the left figure is $\Lambda$.}}
\label{lambdaconstbig}
\end{figure}

The results for larger values of  $ L \Lambda $ are shown  in Figure \ref{lambdaconstbig},  for $\Lambda=1$ fixed and $L=1,2,3,4$. From the figures we see the expected pattern emerging:  by going to larger $L$ at fixed  $\Lambda$  one expects to recover the confined phase of the standard  ${\mathbb C}P^{N-1}$ sigma model,  Eq.~(\ref{ml}).   We indeed see that $\sqrt{\lambda(x)} \to \Lambda$ in the middle of the interval whereas
the condensate   $\sigma$ approaches zero there at the same time. The effects of the boundary remain concentrated near the two extremes  and do not propagate beyond $1/\Lambda$.
\begin{figure}[h!t]
\centerline{
\begin{tabular}{cc}
\epsfxsize=7.5cm \epsfbox{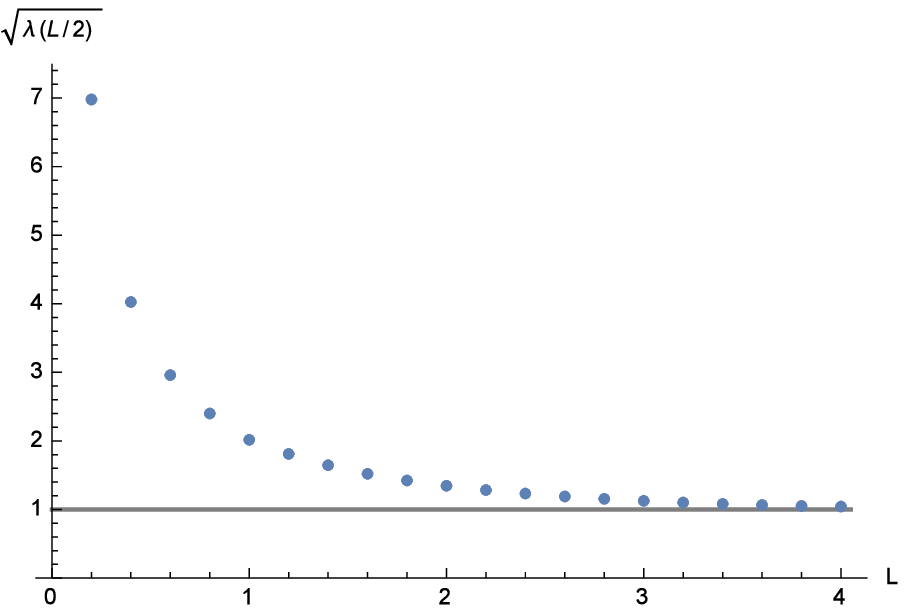}  \ & \ 
\epsfxsize=7.5cm \epsfbox{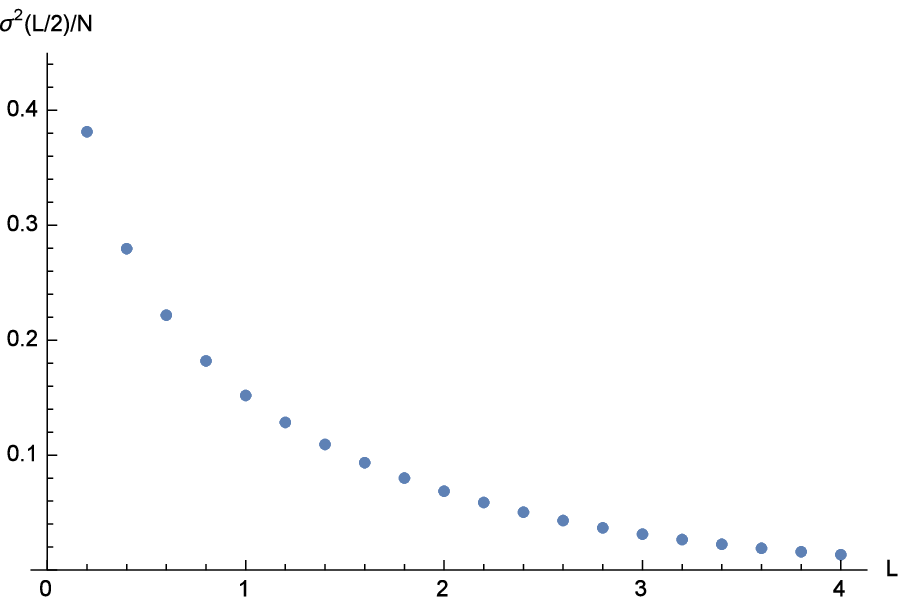} \\
\epsfxsize=7.5cm \epsfbox{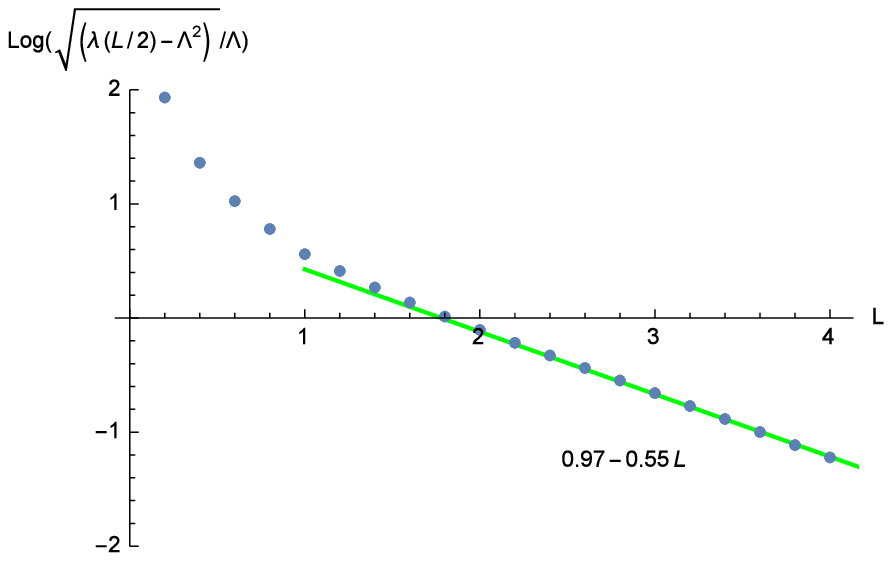}  \ & \ 
\epsfxsize=7.5cm \epsfbox{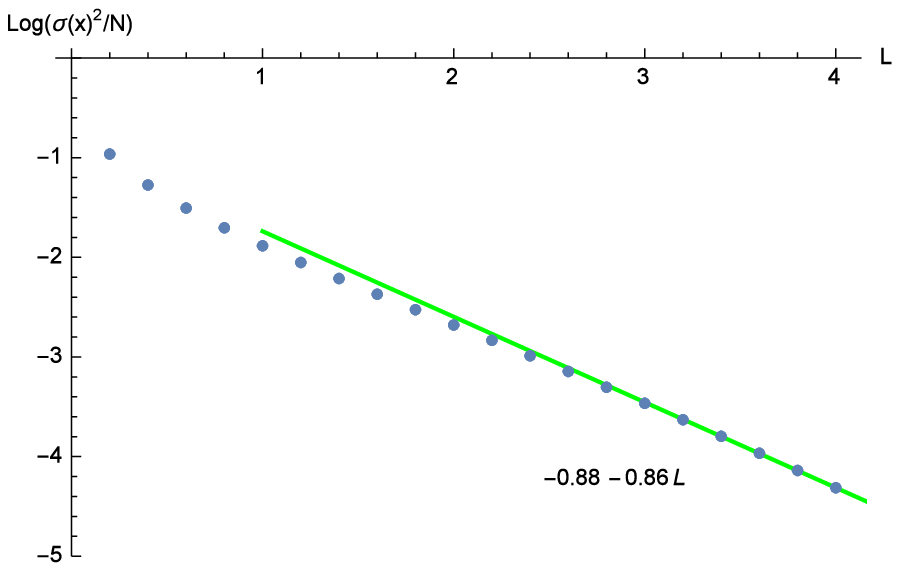} 
\end{tabular}}
\caption{{\footnotesize These plots  are: on the top-left  is $\sqrt{\lambda(L/2)}$,  on the top-right $\sigma(L/2)^2/N$, on the bottom-left $\log{ \left( \sqrt{\lambda(L/2)-\Lambda^2}/\Lambda \right)}$ and on the bottom-right $ \log{\left( \sigma^2(L/2)/N \right) } $.  These plots are obtained for various values of $L$ keeping $\Lambda=1$.   }}
\label{mezzobig}
\end{figure}
\begin{figure}[h!t]
\centerline{
\begin{tabular}{cc}
\epsfxsize=8cm \epsfbox{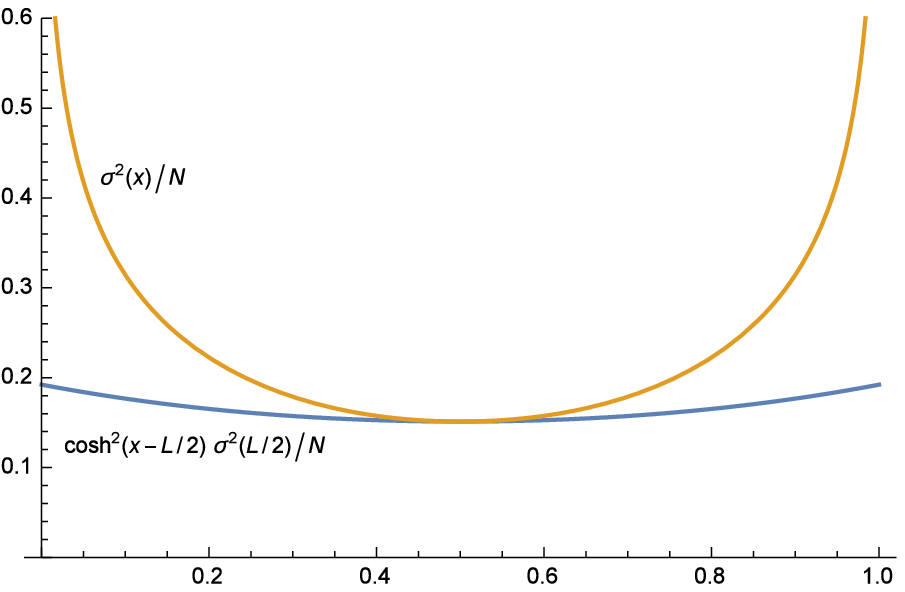}  & 
\epsfxsize=8cm \epsfbox{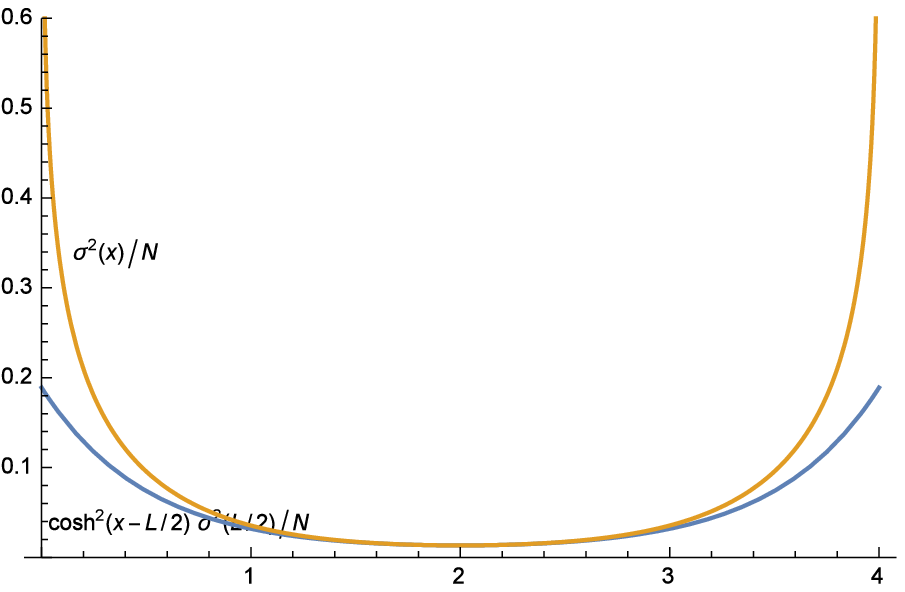} 
\end{tabular}}
\caption{{\footnotesize $\sigma(x)^2/N$ compared with the $cosh$ approximation for $\Lambda=1$ and $L=1$ on the left and $L=4$ on the right. } }
\label{sigmanew}
\end{figure}

\noindent  The values of the fields at the middle point of the interval are given in Figure \ref{mezzobig} where in the bottom line are the corresponding logarithmic plots to show better the rate of convergence to the confinement phase. A linear fit to  the logarithmic plots at large $L$, both for $\log{ \left( \sqrt{\lambda(L/2)-\Lambda^2}/\Lambda \right)}$ and for $ \log{\left( \sigma^2(L/2)/N \right) } $, 
 or 
 \be   \lambda(\tfrac{L}{2})  \simeq  \Lambda^2 +   \Lambda^2 e^{0.97 - 0.55  L}\;;
 \qquad    \sigma(\tfrac{L}{2})^2   \simeq   N  \,   e^{-0.88 - 0.86 L}  \;,     \label{exponentially}
 \ee
  is a fairly  clear signal that the boundary effects are exponentially suppressed  at fixed $x$,  for large values of $L$. 
  
    As a further check of our conclusion, 
Figure \ref{sigmanew} shows  that the field  $\sigma(x)$ becomes indeed  well approximated by a  hyperbolic cosine function in the central region  for large $L $. As seen from Eq.~(\ref{lambda})  
this is what is to be expected for $\sigma(x)$, upon  dynamical generation of the mass for $ n_i $,  $ \brc \lambda \ckt = \Lambda^2$.

\subsection{Absence of the ``Higgs" phase}

The  preceding analysis clearly shows that the solution of the set of the generalized gap equations  is unique:  
it  smoothly approaches in the large $L$ limit    the well-known physics  of the  $CP(N-1)$  model (at large $N$)  in  ``confinement phase" with
\be  \brc \sigma(x) \ckt  =0\;;  \qquad       \brc  \lambda(x) \ckt = m^2= \Lambda^2\;,
\ee
 (See Fig.~\ref{lconst} -  Fig.~\ref{mezzobig}  ). It is however really a simple matter to prove directly the absence of the "Higgs phase", characterized by the condensates
 \be  \brc \sigma(x) \ckt  = {\rm const.} \sim \Lambda \;,\quad       \brc  \lambda(x) \ckt   =   0\;  \label{this}
\ee
for any value of $L$.
The proof is basically identical to the proof already given (Subsection~\ref{sec:proof}) of the fact that translationally invariant condensates, $\lambda$ and $\sigma$ constant, are not consistent with 
the generalized gap equations: rather, it is a special case of it.  Indeed, if the $n$ field mass is not generated, $\lambda \equiv 0$, 
the approximation used for higher levels ($n \gg 1$)   (see Eq.~(\ref{largen}), Eq.~(\ref{massless}))   is actually valid for all levels.  Thus one finds 
that  the first term of  the gap equation (\ref{lnonconst}) is given exactly by,
\beq
\frac{N}{2} \, \sum_n\frac{1}{\omega_n}\left( f_n(x)^2 -\frac1L \right) =   \frac{N}{2 \pi} \log
\left(2 \sin{\left(\frac{\pi x}{L} \right)}  \right) \ ,  \label{masslessbis}
\eeq
which is certainly not equal to $0$ identically.  This proves that the Higgs phase is not possible in the presence of the boundaries, except for the case with the periodic boundary conditions.  

\section{Discussion}

In this paper we have studied in detail the two-dimensional  ${\mathbb C}P^{N-1}$ model with a finite space interval.  In other words the system is defined on a
finite-width worldstrip, in which the fields propagate. We find that the system has a unique phase   which smoothly approaches  in the large $L$ limit  the standard $2D$    ${\mathbb C}P^{N-1}$ system in confinement phase, with (constant) mass generation  for the $n_i$ fields, which are then confined by a $2D$  Coulomb force. 

This kind of model is of great interest in physics. For instance, it appears as the low-energy effective theory describing the quantum excitations of monopole-vortex soliton complex \cite{MVComplex}. In such a context,  the ${\mathbb C}P^{N-1}$  model  describes the nonAbelian orientational zeromodes of the nonAbelian vortex (string),  
whereas  its boundaries represent the monopoles arising from a higher-scale gauge-symmetry breaking, carrying the same orientational ${\mathbb C}P^{N-1}$ 
moduli. NonAbelian monopoles, which are not plagued by the well-known pathologies, emerge in such a context, and the infrared properties of 
such a soliton complex is an important question, defining the dynamical properties of the nonAbelian monopoles. 
 
As the  ${\mathbb C}P^{N-1}$ sigma model in two dimensions  is asymptotically free, i.e., it is a theory which is scale invariant in the UV, one might wonder what the meaning 
of the "finite interval" $L$ is.  The point is that the behavior of the system in the UV - where the effects of the space boundaries should be negligible -  contains in itself by dimensional transmutation the renormalization-invariant 
mass scale $\Lambda$ (equivalent to specifying the value of the coupling constant at the ultraviolet  cutoff, $\Lambda_{UV}$), whether or not the system dynamically 
flows actually  into the infrared regime at $1/\Lambda$ where the interactions become strong.  
The space interval $L$  can be defined in reference to $1/\Lambda$. 

But this  implies  that for $L \ll 1/ \Lambda $ quantum fluctuations would not become strongly coupled,  their wavelengths 
constrained to be  smaller than $1/\Lambda$.  In other words, the system effectively reduces to a finite dimensional object propagating in time, i.e.,  quantum mechanics,  
rather than a proper $2D$ quantum field theory.  This would explain our finding that  the ${\mathbb C}P^{N-1}$ model with a space boundary has a unique phase. 

With the periodic boundary condition, the story could be different. The translational invariance is unbroken, and it could be a property of the vacuum: the constant mass   $\lambda(x)=m^2$  	
(or $\sigma$) generation of the order of $\Lambda$  is certainly consistent with the functional saddle-point equations, as we have shown.  
  Even $L< 1/\Lambda$ the system thus maintains $2D$ field-theoretic properties,  and it could be  \cite{Monin:2015xwa} that,  at large $N$,  the system shows  a phase transition at a critical length $L_{crit}  =  O(1/ \Lambda)$,  from  confinement  ($m \sim \Lambda,\, \sigma=0$)   to Higgs   ($\sigma\sim \Lambda,m=0$) phase.

\section*{Acknowledgments}
 We thank Alessandro Betti and Simone Giacomelli for discussions, Ettore Vicari for useful comments,  and Gerald Dunne for communications.  The  work of SB is funded by the grant ``Rientro dei Cervelli Rita Levi Montalcini'' of the Italian government. The present research work is supported by the INFN special research project grant, GAST ("Gauge and String Theories").

\appendix

\section{Eigenfunctions near the boundaries}
\label{eigen}

$\lambda(x)$ satisfying the generalized gap equation was found to 
behave  as  
\begin{eqnarray}
 \lambda(x)\approx -\frac1{2x^2\log x} \approx \frac{\Lambda_0^2}2
  e^{\frac{2}y} y\;,     \label{lambda0}
\end{eqnarray}
around $x=0$,  
where 
\beq   y=y(x)\equiv \frac1{-\log(\Lambda_0 x)}\;, \qquad  x= \Lambda_0^{-1} \,   e^{-1/y}  \;, 
\eeq
and
 $\Lambda_0$ is some constant.  In terms of the new coordinate $y$,
$\tilde \lambda(y)\equiv \lambda(x)$
has an essential singularity at $y=0$.
It is natural to assume that $\lambda(x)$ can be expanded around the boundary at
$x=0  \, (y=0)$ as 
\begin{eqnarray}
 \lambda(x)=\tilde \lambda(y)
=\Lambda_0^2 \sum_{n=0}^\infty  e^{-\frac{n-2}y}
  v_n(y)\;,  \label{eq:lambdaexp}
\end{eqnarray}
where $v_n(y)$ are functions with no  essential singularities.
 Especially, we set  for   $ v_0(y)$
\begin{eqnarray}
 v_0(y)=\frac{y}2+ \sum_{n=2}^\infty a_n y^n\;.
\end{eqnarray}
The $n=0$ term in  (\ref{eq:lambdaexp})  gives the  dominant contribution  since  
\begin{eqnarray}
 \lim_{y\to + 0} \frac{e^{-\frac{n-1}y}v_{n+1}(y)}{e^{-\frac{n-2}y} v_n(y)}=0\;.
\end{eqnarray} 
Note that the knowledge of an exact form of $v_0(y)$ and poles of $v_1(y)$
is needed to determine the divergent contribution to the following integral  
\begin{eqnarray}
 r \int_\epsilon^{L-\epsilon}\lambda(x)dx 
\approx 2r\int_\epsilon \left(\frac{v_0(y(x))}{x^2}+\frac{v_1(y(x))}{x}\right)dx= \frac{N}{2\pi \epsilon
  }+{\cal O}\left(-\frac{1}{\epsilon \log\epsilon }\right)
\end{eqnarray}
and to renormalize the total energy.    
An incomplete knowledge of the subdominant divergences in  $\lambda(x)$ and $\sigma(x)$  for the moment 
prevents us from removing these divergences from the energy unambiguously.  
We expect that these divergences are independent of the length $L$; the resulting finite contribution may be interpreted as the
mass of the endpoints (the monopole mass).   

Given the potential $\lambda(x)$, the 
behavior of an eigenfunction $f(x)$ near the boundaries can be inferred directly from    
\begin{eqnarray}
 \left(-\partial_x^2+\lambda(x)\right)f(x)=\omega^2 f(x)\;,\label{eq:feq}
\end{eqnarray}
where  we dropped the level number,  as  the leading behavior near $x=0$ or $x=L$ turns out to be independent of the energy level. 
We assume an expansion of the form
\begin{eqnarray}
 f(x)= \sum_{n=0}^\infty e^{-\frac{n+\rho}y} h_n(y)\;,\qquad 
h_n(y)=\sum_{k=0}^\infty c_{n,k} \, y^{k+\delta_n}\;.\label{eq:fexp}
\end{eqnarray}
for $f(x)$,  with constants $\rho, \delta_n$. 
Substituting   (\ref{eq:lambdaexp}) and (\ref{eq:fexp}) into 
Eq.~(\ref{eq:feq}) one sees that the dominant terms on the left hand side are  of the order of 
$O(e^{\frac{2-\rho}y})$, whereas the right hand side is $O(e^{\frac{-\rho}y})$, which is smaller by
$e^{-2/y} \sim x^2$.  Requiring thus  that the leading 
$e^{\frac{2-\rho}y}$  terms  of  Eq.~(\ref{eq:feq})   to vanish, one is led to  the
following equation 
\begin{eqnarray}
 y^4 h_0''(y)+y^2(2y+2\rho-1)h_0'(y)+(\rho(\rho-1)-v_0(y))h_0(y)=0\;.  
\label{eq:h0}
\end{eqnarray}
By inserting the expansion (\ref{eq:fexp}) for $h_0(y)$,    this can be solved 
 recursively  for $c_{0,k}$:  
\begin{eqnarray}
\rho(\rho-1) \, c_{0,k+2}+\left((k+\delta_0+1)(2\rho -1)-\frac12
		       \right)c_{0,k+1}
\nonumber\\
=\left( a_2-(k+\delta_0)(k+\delta_0+1) \right)
c_{0,k}+\sum_{m= 1}^\infty a_{m+2}\,  c_{0,k-m}\;.
\end{eqnarray}
As   $c_{0,k}=0  \,\,  (k<0)$,   these recursion relations require  
\begin{eqnarray}
 (\rho, \delta_0)= \left(0,-\frac12\right),\quad  {\rm or~} \qquad  
\left(1,\frac12\right)\;,
\end{eqnarray}
that is,  
\be     f(x) \sim        \sqrt{-\log x}\;,   \qquad  {\rm or} \qquad 
  \frac{x}{\sqrt{-\log x}} \;,  
\ee
near $x=0$.  Actually the above recursive relations define two different solutions
of the equation (\ref{eq:h0})  for $h_0(y)$. 

Similarly, 
by considering terms of the order of   $\exp{\frac{-n+2- \rho}y}$, 
one finds a recursive relation for $h_n(y)$
which has a {\it unique} solution,   {\it  given}
 $\{h_0(y),h_1(y),\cdots,h_{n-1}(y), \omega\}$.

This way 
one finds the  general solution of Eq.~(\ref{eq:feq}):  
\begin{eqnarray}
 f(x)=A_1 f^{(1)}_{\omega}(x)+A_2 f^{(2)}_{\omega}(x)\;,  
\end{eqnarray}
where $f^{(1,2)}(x)$ behave around the boundary at $x=0$ as
\begin{eqnarray}
 f^{(1)}_\omega (x)\approx \sqrt{-\log x}\;,\qquad   f^{(2)}_{\omega}(x) \approx
  \frac{x}{\sqrt{-\log x}} \;.
\end{eqnarray}
As  $f^{(1)}_{\omega}(x), f^{(2)}_{\omega}(x)$ yield  a nonvanishing Wronskian
\begin{eqnarray}
 \lim_{x\to 0}(f^{(1)}_{\omega}(x)f^{(2){}'}_{\omega}(x)
- f^{(1){}'}_{\omega}(x)f^{(2)}_{\omega}(x))=1\;, 
\end{eqnarray} 
they clearly  represent two  linearly independent solutions for given $\omega$. 

Our boundary condition requires that the functions $f_n(x)$ describing the quantum fluctuations of the $n_i$ fields choose 
\be  f_n(x) \propto   f^{(2)}_{\omega}(x) \approx
  \frac{x}{\sqrt{-\log x}} \;,
\ee
which is the result reported in  (\ref{result}). On the other hand, the classical field $\sigma$, corresponding to the zero  ($\omega=0$) mode,   was found to opt for (see Eq.~(\ref{sigmadiv}))
\be   \sigma = \sqrt{\frac{N}{2\pi}} f^{(1)}_\omega (x)\approx   \sqrt{\frac{N}{2\pi}} \sqrt{-\log x}\; .
\ee

\end{document}